\documentclass [11pt,a4paper] {article}

\usepackage[cp1252]{inputenc}
\usepackage[english]{babel}
\usepackage{amssymb}
\usepackage{amsmath}
\usepackage{amsfonts,amssymb}
\usepackage[dvips]{graphicx}

\newtheorem{theorem}{Theorem}
\newtheorem{corollary}{Corollary}

\begin{document}

\title{Any realistic model of a physical system must be computationally realistic}

\author{Arkady Bolotin\footnote{$Email: arkadyv@bgu.ac.il$} \\ \textit{Ben-Gurion University of the Negev, Beersheba 84105, Israel}}

\maketitle

\begin{abstract}
\noindent It is argued that any possible definition of a realistic physics theory -- i.e., a mathematical model representing the real world -- cannot be considered comprehensive unless it is supplemented with requirement of being computationally realistic. That is, the mathematical structure of a realistic model of a physical system must allow the collection of all the system's physical quantities to compute all possible measurement outcomes on some computational device not only in an unambiguous way but also in a reasonable amount of time.

\indent In the paper, it is shown that a deterministic quantum model of a microscopic system evolving in isolation should be regarded as realistic since the NP-hard problem of finding the exact solution to the Schr\"odinger equation for an arbitrary physical system can be surely solved in a reasonable amount of time in the case, in which the system has just a small number of degrees of freedom. In contrast to this, the deterministic quantum model of a truly macroscopic object ought to be considered as non-realistic since in a world of limited computational resources the intractable problem possessing that enormous amount of degrees of freedom would be the same as mere unsolvable.
\end{abstract}

\section{Introduction}

\noindent Recently a definition of realistic physics theories -- i.e., mathematical models representing the real world -- has proposed, according to which a model of a physical system is classified as realistic if and only if, according to the mathematical structure of this model, the collection of all physical quantities written in the system \textit{unambiguously} determines the probabilities of all possible measurement outcomes \cite{Gisin}. Attractive and credible as this definition might appear, it nevertheless fails to mention that before anything else a realistic model has to be \textit{computationally} realistic. This means that the mathematical structure of a realistic model of a physical system must allow the collection of all the system physical quantities to determine all possible measurement outcomes (or their probabilities) not only in an unambiguous way but also in a realistic -- that is, \textit{reasonable} -- amount of time. Undeniably, the defining characteristic of any mathematical physics model is that it makes falsifiable or testable predictions. So, if the mathematical model were insoluble in a reasonable amount of time (even with access to a supercomputer), then the model would not have a realistically testable predictive content, and therefore the term ``realistic" would be hardly applicable to it.\\

\noindent Of course, the requirement of being computationally realistic would be completely redundant for the definition of a realistic physics theory if every physics model were capable of being solvable in reasonable time as soon as enough computational resources were thrown at the model. The goal of this paper is to show that, in all likelihood, such is not the actual state of things in \textit{quantum theory}. In particular, it will be shown that the deterministic quantum model of a microscopic system evolving in isolation should be regarded as a realistic model, whilst the quantum model of a truly macroscopic object (built around exact solutions to the object's Schr\"odinger equation) ought to be considered as non-realistic.

\section{Schr\"odinger's equation computational complexity}

\noindent To demonstrate this, the complexity of Schr\"odinger's equation, which is one of the fundamental equations of quantum theory that describes how the quantum state of a physical
system changes with time, should be analyzed.

\begin{theorem}

Let $H$ be the Schr\"odinger Hamiltonian for a physical system, the vector $\!\left.\left|\psi \!\left(t\right)\!\right.\right\rangle$ be the state of this system at time $t$, the wave function $\Psi\!\left({\mathbf r},t\right)$ be the product $\!\left\langle {\mathbf r}\left|\psi \!\left(t\right)\!\right.\right\rangle$, where ${\mathbf r}\!=\!\left({{\mathbf r}}_1,{{\mathbf r}}_2,\dots ,{{\mathbf r}}_N\right)$ denotes the sets of position vectors such that ${{\mathbf r}}_n\!=\!\left({x}_{n1},{x}_{n2},{x}_{n3}\right)$ and $N$ is the system constituent particle number. Then the runtime complexity of the operations needed to verify that any given solution $\Psi\!\left({\mathbf r},t\right)$ is indeed the solution to the Schr\"odinger equation $i\hbar\:{\partial\Psi\!\left({\mathbf r},t\right)}/{\partial t}\!=\!H\Psi\!\left({\mathbf r},t\right)$ is polynomially bounded.

\end{theorem}

\noindent As it can be readily seen, to calculate the complexity of verification (defined as value $L$) it is sufficient to find the minimal number of elementary operations required to compute the effects of the Schrödinger Hamiltonian $H$ on the solution $\Psi\!\left({\mathbf r},t\right)$. After been projected into the position basis $\{\!\left.\left|{\mathbf r}\!\right.\right\rangle\}$ the generic form of the Schrödinger Hamiltonian $H$ can be expressed as the following sum:

\begin{equation} \label{1}
   H =-\frac{{\hbar}^2}{2}\sum^N_{n=1}{\frac{1}{m_n}\frac{{\partial }^2 }{\partial {{\mathbf r}}^2_n}}-\frac{{\hbar}^2}{2M}\sum^N_{j\ne l}{\frac{\partial}{\partial {{\mathbf r}}_j}\frac{\partial}{\partial {{\mathbf r}}_l}}+{\widehat{{\rm H}}}_{{\rm s}}+V\!\left({{\mathbf r}}_1,{{\mathbf r}}_2,\dots ,{{\mathbf r}}_N,t\right) \;\;\;\; , 
\end{equation}
\smallskip

\noindent where $m_n$ is the mass of the system constituent particle $n$, $M$ denotes the mass of the collection of the constituent particles resulting in the extra kinetic energy $K$ (owing to the dependency $K$ on the spatial configuration of the particles), $\widehat{{\rm H}}_{\rm s}$ is the term accounting for the presence of the constituent particles' spins (this terms may include spin-orbit coupling, spin-rotation coupling and spin-spin coupling). Additionally, in the case of the external electromagnetic field existing represented by the vector potential ${{\mathbf A}}\!=\!\left(A_1,A_2,A_3\right)$, the interaction of the constituent particle of charge $q_n$ with the electromagnetic field can be taken into account by replacing the partial derivatives ${\partial }\!/{\partial {x}_{nj}}$ by  ${\partial }\!/{\partial {x}_{nj}}-{iq_nA_j}\!/{\hbar c}$. So, making use of the results of the papers \cite{Paterson,Baur}, the complexity $L$ can be presented as follows:

\begin{equation} \label{2} 
   L\!\left(
        H\Psi 
   \!\right)
   =
   L\!\left(
         \frac{{\partial }^2\Psi }{\partial {{\mathbf r}}^2_1},\dots ,\frac{{\partial }^2\Psi }{\partial {{\mathbf r}}^2_N},\frac{{\partial }\Psi }{\partial {{\mathbf r}}_1},\dots ,\frac{{\partial }\Psi }{\partial {{\mathbf r}}_N}
   \right)
   \le
   O\!\left(\!{N^2}\!\right)
   \cdot
   {\mit cost}\!\left(
                    \Psi\!\left({\mathbf r},t\right)
            \right)
\;\;\;\;  ,
\end{equation}
\smallskip

\noindent where only nonscalar arithmetic operations (i.e., binary operations whose both operands involve the solution $\Psi$) are considered contributed to the complexity of verification $L$, whereas additions/subtractions and multiplications by arbitrary scalars $k\!\left({\mathbf r},t\right)$ are allowed for free, ${\mit cost}\!\left(\Psi\!\left({\mathbf r},t\right)\right)$ denotes the computational cost of the wave function evaluation at particular numerical values ${\mathbf r}\!=\!\left({{\mathbf r}}_1,{{\mathbf r}}_2,\dots ,{{\mathbf r}}_N\right)$ and $t$. In order to the interpretation of solutions to Schr\"odinger's  equation to make sense, it must be \textit{feasible} to evaluate the given solution $\Psi\!\left({\mathbf r},t\right)$ at arbitrary $\mathbf r$  and $t$ because otherwise it would be impossible to use $\Psi\!\left({\mathbf r},t\right)$ to compute a measurable observable of the quantum system. In conformity with Cobham's thesis \cite{Cobham}, the function $\Psi\!\left({\mathbf r},t\right)$ can be feasibly evaluated only if it can be evaluated on some computational device in polynomial time. This implies that ${\mit cost}\!\left(\Psi\!\left({\mathbf r},t\right)\right)\!=\!{poly\left(N\right)}$, and hence, the verification complexity $L$ is upper-bounded by a polynomial. $\square$

\begin{corollary}
The decision problem of Schr\"odinger's equation  ``Does the ground state of a physical system $\Psi_g\!\left({\mathbf r}\right)$ have energy $E_g\le0$?'' is in the NP complexity class of computational problems, whose solutions can be verified in polynomial time.
\end{corollary}

\noindent Certainly, if the ground state $\Psi_g\!\left({\mathbf r}\right)$ with the energy $E_g\le0$ is given, then by substituting it back into Schr\"odinger's equation one can prove that $\Psi_g\!\left({\mathbf r}\right)$ is indeed the solution to this equation in only a polynomial amount of time.

\begin{theorem}
Given an exact generic algorithm capable of solving exactly the Schr\"odinger equation for an arbitrary physical Hamiltonian (that is, for any and all possible physical systems), any problem in the NP complexity class can be solved with only polynomially more work.
\end{theorem}

\noindent Let $A\!\left(\!{\Psi}\!\right)$ be such an algorithm. As $A\!\left(\!{\Psi}\!\right)$ can solve exactly the Schr\"odinger equation for all Hamiltonians, it can also solve the Schrödinger equation for Ising Hamiltonians that describe Ising models of a spin glass \cite{Fischer,Guerra}. On the other hand, the paper \cite{Lucas} explains how ``all the famous NP problems'' (such as Karp's 21 NP-complete problems \cite{Garey}) can be written down as Ising models with only a polynomial number of steps (to be exact, with a polynomial number of spins which scales no faster than $N^3$). Therefore, in just a polynomial number of steps one can get from any NP-complete problem to the Schr\"odinger equation for the Hamiltonian of an Ising spin glass, whose decision problem ``Does the ground state of a glass $\Psi_g\!\left({\mathbf r}\right)$ have energy $E_g\le0$?'' solves the NP-complete problem of interest using the algorithm $A\!\left(\!{\Psi}\!\right)$. From here, the proof of the theorem follows: Since an arbitrary NP problem is polynomial-time reducible to any NP-complete problem and thus to the decision form of the Ising model, any problem in NP can be solved exactly by the algorithm $A\!\left(\!{\Psi}\!\right)$ with only polynomially more work. $\square$

\begin{corollary}
The problem of solving exactly Schr\"odinger's equation for an arbitrary physical system is NP-hard.
\end{corollary}

\noindent Seeing as the algorithm $A\!\left(\!{\Psi}\!\right)$ that solves this problem can also solve any problem in the NP complexity class, the given problem is NP-hard.

\begin{corollary}
If the algorithm $A\!\left(\!{\Psi}\!\right)$ were efficient, the class NP would be equal to the class P of computational problems solvable in polynomial time. If P$\ne$NP, the algorithm $A\!\left(\!{\Psi}\!\right)$ could not be efficient, i.e., its runtime complexity could not be polynomial in $N$.
\end{corollary}

\begin{theorem}
If P$\ne$NP, Schr\"odinger's equation for a truly macroscopic object would not have an exact solution reachable in polynomial time.
\end{theorem}

\noindent As an entity of classical realm, a truly macroscopic object contains a large (and essentially unchecked)
 number of constituent microscopic particles that interact constantly with many different physical systems (of various properties and scales) within the vast causal horizon for the object. As such, its Hamiltonian must be a sum of (practically) all possible physical Hamiltonians, and hence an algorithm capable of solving exactly the Schr\"odinger equation for this object must be the exact generic algorithm $A\!\left(\!{\Psi}\!\right)$. Therefore, if P$\ne$NP, Schr\"odinger's equation for a truly macroscopic object could not be solved exactly in polynomial time. $\square$ \\

\noindent Assuming that P$\ne$NP, is it possible that the exact generic algorithm $A\!\left(\!{\Psi}\!\right)$ is significantly faster than exhaustive search (an exact generic algorithm as well), which, according to the postulates of quantum mechanics, runs exponentially in $N$ as $O({2^N})$?

\begin{theorem}
In all probability, if P$\ne$NP, then the runtime complexity of the exact generic algorithm $A\!\left(\!{\Psi}\!\right)$  could not be sub-exponential in $N$.
\end{theorem}

\noindent Suppose that $A\!\left(\!{\Psi}\!\right)$ is a sub-exponential time algorithm. Since any NP-complete problem -- including the 3-SAT problem -- can be written down as the Schr\"odinger equation for the Hamiltonian of an Ising spin glass, it follows that the algorithm $A\!\left(\!{\Psi}\!\right)$ can solve any NP-complete problem in sub-exponential time. However, according to the widely believed conjecture (called \textit{the exponential time hypothesis}), the 3-SAT problem does not have a sub-exponential time algorithm \cite{Impagliazzo,Woeginger,Lokshtanov}. Hence, if the runtime complexity of $A\!\left(\!{\Psi}\!\right)$ were sub-exponential in $N$, then the exponential time hypothesis could be shown to be false. $\square$

\begin{corollary}
In all probability, if P$\ne$NP, then an algorithm solving exactly Schr\"odinger's equation for a truly macroscopic object could not be significantly faster than exhaustive search.
\end{corollary}

\section{Discussion}

\noindent It is clear that (even if P$\ne$NP) the NP-hard problem of solving exactly Schr\"odinger's equation for an arbitrary physical system could surely be solved in a reasonable amount of time for the instance where the system is composed of just a few constituent microscopic particles. This means that the quantum model of a microscopic system evolving in isolation should be regarded as a realistic model. In contrast to this, the quantum model of a truly macroscopic object (such as a macroscopic detector, Schr\"odinger's cat, an observer), ought to be considered as a non-realistic model. Indeed, as it follows from Corollary 4, it is quite likely that if P$\ne$NP, then to everyone living in a world of limited in time computational resources, the problem of solving exactly Schr\"odinger's equation for a truly macroscopic object is the same as a mere unsolvable problem. \\

\noindent On the other hand, if the macroscopic object is considered as a system with only a few controlled or measured degrees of freedom (among many others that are uncontrolled or unmeasured) then the object's quantum model (an inexact, probabilistic one as it can only provide an incomplete description of the system) can surely be solved in reasonable time and thus regarded as a realistic physics model.


\begin{thebibliography}{11}

\bibitem{Gisin}\label{Gisin}
Gisin N 2014 A possible definition of a Realistic Physics Theory \textit{Preprint} arXiv:1401.0419 [quant-ph]

\bibitem{Paterson}\label{Paterson}
Paterson M and Stockmeyer L 1973 On the number of nonscalar multiplications necessary to evaluate polynomials \textit{SIAM J. Comput.} \textbf{2} 1

\bibitem{Baur}\label{Baur}
Baur W and Strassen V 1983 The complexity of partial derivatives \textit{Theoretical Computer Science} \textbf{22} 317-330

\bibitem{Cobham}\label{Cobham}
Cobham A 1964 The intrinsic computational difficulty of functions \textit{Proc. Int. Cong. for Logic, Methodology and Philosophy of Science II} (North Holland)

\bibitem{Fischer}\label{Fischer}
Fischer K and Hertz J 1991 \textit{Spin Glasses} (Cambridge University Press)

\bibitem{Guerra}\label{Guerra}
Guerra F and Toninelli F 2002 The thermodynamic limit in mean field spin glass models \textit{Communications in Math. Phys.} 230 \textbf{1} 71-79 (\textit{Preprint} arXiv:0204280 [cond-mat])

\bibitem{Lucas}\label{Lucas}
Lucas A 2014 Ising formulations of many NP problems \textit{Frontiers in Physics} \textbf{2} 5 (\textit{Preprint} arXiv:1302.5843 [cond-mat.stat-mech])

\bibitem{Garey}\label{Garey}
Garey M and Johnson D 1979 \textit{Computers and Intractability: a Guide to the Theory of NP-Completeness} (New York: Freeman \& Co)

\bibitem{Impagliazzo}\label{Impagliazzo}
Impagliazzo R, Paturi R, and Zane F 2001 Which problems have strongly exponential complexity? \textit{J. Comput. System Sci.} \textbf{63} 512-530

\bibitem{Woeginger}\label{Woeginger}
Woeginger G 2003 \textit{Exact Algorithms for NP-hard Problems: A Survey} Combinatorial Optimization - Eureka, You Shrink! (Springer-Verlag) pp. 185-207

\bibitem{Lokshtanov}\label{Lokshtanov}
Lokshtanov D, Marx D, Saurabh S 2011 Lower bounds based on the exponential time hypothesis \textit{Bulletin of the EATCS} \textbf{84} 41-71

\end{thebibliography}
\end{document}